\documentclass[12pt]{article}
\usepackage{amsmath,amssymb}

\usepackage{amsfonts,amsthm,amsmath,amssymb,graphicx}

 \def\frac#1#2{{#1\over #2}}

\def\be{\begin{equation}}
\def\ee{\end{equation}}
\def\ba{\begin{eqnarray}}
\def\ea{\end{eqnarray}}

 \def\f {\frac}

 \def\no{\nonumber \\}

 \def\la{\langle}
 \def\lb{\rangle}

\begin{document}

\begin{titlepage}
\thispagestyle{empty}

\begin{flushright}
YITP-16-84
\\
\end{flushright}

\begin{center}
\noindent{{\Large \textbf{Butterflies from Information Metric}}}\\
\vspace{2cm}
{ Masamichi Miyaji}

\vspace{1cm}
{\it Yukawa Institute for Theoretical Physics,\\
Kyoto University, Kyoto 606-8502, Japan}
\vskip 2em
\end{center}

\begin{abstract}
We study time evolution of distance between thermal states excited by local operators, with different external couplings. We find that growth of the distance implies growth of commutators of operators, signifying the local excitations are scrambled.
We confirm this growth of distance by holographic computation, by evaluating volume of codimension 1 extremal volume surface.
We find that the distance increases exponentially as $e^{\f{2\pi t}{\beta}}$.
Our result implies that, in chaotic system, trajectories of excited thermal states exhibit high sensitivity to perturbation to the Hamiltonian, and the distance between them will be significant at the scrambling time. We also confirm the decay of two point function of holographic Wilson loops on thermofield double state.
 \end{abstract}

\end{titlepage}
\newpage

\section{Introduction}
The study of chaos in terms of AdS/CFT correspondence  \cite{Maldacena:1997re} has revealed new perspectives on quantum field theory and gravity. Its relation to black hole information paradox \cite{Hayden:2007cs} \cite{Kitaev} and Fast scrambling conjecture \cite{Sekino:2008he}\cite{Shenker:2013pqa}\cite{Maldacena:2015waa} is of particular interest. \\

 In this article, we study scrambling of local excitation in chaotic system. Consider a thermal system at equilibrium excited by a local operator $W$. Sometime after the excitation, the excitation will lose its identity, so that it can not be distinguished from other excitations by local measurement. We can distinguish them only by measuring total system,
  so that the excitation is scrambled. One way to define indistinguishability is to measure the growth of complexity of the excitation $W(t)$. When the perturbation $W(-t_w)$ is scrambled, then it should be a complicated sum of variety of local operators. Complexity of operators can be measured by the growth of expectation value of square of commutator 

\be [W(t),~V]\ee\\
 for all local $V$ with $[W,~V]=0$. The expectation value is small at early time, and grows as $e^{\lambda_L t}$ at late time, where $\lambda_L$ is called Lyapunov exponent. Such growth of commutators was already considered in \cite{Larkin}, and was interpreted as diagnosis of chaos, because semiclassically the commutator $\langle [q(t),~p(0)][q(t),~p(0)]^{\dagger}\rangle$ is equal to the square of Poisson bracket 

\be\{ q(t),~p(0)\}_{P}^2=(\f{\delta q(t)}{\delta q(0)})^2,\ee\\
whose exponential growth implies high sensitivity of classical trajectories to initial configurations. 
$\lambda_{L}$ is known to obey a bound $\lambda\leq \f{2\pi}{\beta}$ \cite{Maldacena:2015waa} in large N theory. The bound is saturated by the holographic CFT dual to classical Einstein gravity, and the dual geometry 
can be approximated by shock wave geometry \cite{Shenker:2013pqa}\cite{Shenker:2013yza}\cite{Shenker:2014cwa}\cite{Roberts:2014isa}\cite{Roberts:2014ifa}\cite{Caputa:2015waa}. Lyapunov exponents of various theories are computed in \cite{Kitaev1}\cite{Stanford:2015owe}\cite{Michel:2016kwn}\cite{Caputa:2016tgt}\cite{Gu:2016hoy}\cite{Fitzpatrick:2016thx}\cite{Chen:2016cms}. Related studies include \cite{Jackson:2014nla}\cite{Polchinski:2015cea}\cite{Hosur:2015ylk}\cite{Gur-Ari:2015rcq}\cite{Berkowitz:2016znt}\cite{Swingle:2016var}\cite{Perlmutter:2016pkf}\cite{Sircar}.\\

 In this paper, we will characterize scrambling of operator by a quantum information theoretic quantity called 
Fisher information metric and quantum fidelity. Quantum fidelity is a measure of similarity between two quantum states.
It is defined by square of absolute value of inner product of two states 

\be
|\langle\psi_{\lambda+\delta\lambda}|\psi_{\lambda}\rangle|^2.
\ee\\
 Decay of quantum fidelity implies 
the distance between states becomes larger. We call the infinitesimal part of the fidelity $G_{\lambda\lambda}$ as Fisher information metric,

\be
|\langle\psi_{\lambda+\delta\lambda}|\psi_{\lambda}\rangle|^2
=1-2G_{\lambda\lambda}(\delta\lambda)^2+\mathcal{O}((\delta\lambda)^3).
\ee\\
Fisher information metric is a metric on the manifold of quantum states, parametrized by parameter $\lambda$.
By definition, Fisher information metric is positive and is covariant quantity. 
In the context of AdS/CFT,  the holographic dual of
 inner product of states with different marginal couplings was proposed in \cite{MIyaji:2015mia},
 and it was argued that volume of codimension 1 surface with extremal volume
  in dual geometry is proportional to the Fisher information metric.\\
  
  Quantum fidelity has been a useful tool to study chaos in quantum systems \cite{Peres}\cite{Jalabert:2001}
  \cite{Jacquod}\cite{Cerruti}\cite{Prosen1}\cite{Karkuszewski}\cite{Prosen2}\cite{Emerson}\cite{Gorin}\cite{Goussev}.
  When two slightly different states evolve with identical Hamiltonians, the fidelity between two states
   is conserved because of linearity and unitarity of quantum mechanics, so "butterfly effect" does not appear.
    Nonetheless, butterfly effect plays role when we evolve two states
    with slightly different Hamiltonians $H_{\lambda}=H_0+\lambda V$ and $H_{\lambda+\delta\lambda}=H_0+(\lambda+\delta\lambda) V$. $\lambda$ represents a coupling of external environment or internal imperfections. In this case inner product 
     
     \be
     |\langle\psi_{\lambda+\delta\lambda}|e^{iH_{\lambda+\delta\lambda}t}e^{-iH_{\lambda}t}|\psi_{\lambda}\rangle|
     \ee\\     
     is not generically conserved, and it decays rapidly if either $H_{\lambda}$ or $H_{\lambda+\delta\lambda}$ is chaotic.
      The decay of fidelity signifies difficulty of revival of quantum state by imperfect time reversal procedure,
 and is also a measure of strength of decoherence.\\
    
In this article, we consider distance between thermal states excited by local operator with slightly different
 Hamiltonians. Assuming the difference between Hamiltonians is infinitesimally small, we find that the distance is proportional to expectation values of commutators of local operators. This implies the rapid growth of distance signifies the excitation is scrambled.
In particular, in large N chaotic theory, we show that growth of Fisher information metric is proportional to
 $e^{\f{2\pi}{\beta}t}$, from holographic computation \cite{MIyaji:2015mia}. Intuitive interpretation of our result is that the trajectories of thermal states excited by local operators are highly sensitive to external environment or internal imperfections, so it diagnoses butterfly effect. In addition, we study two point functions between Wilson loop operators on different boundaries. We confirm that it decays rapidly.\\
  
  {\bf Note}: The volume of maximal volume surface in shock wave geometry was computed up to $\mathcal{O}(1)$ constant
   in \cite{Stanfords:2014cas}, in order to confirm local perturbation does not change complexity significantly. Our study is focused on that $\mathcal{O}(1)$ constant. Also, similarity between timescales in scrambling and the decay of Loschmidt echo was pointed out in \cite{Swingle:2016var}. Here we provide a direct connection between them and give a concrete example by holographic computation.
   
\section{Fisher information metric}
In this section, we introduce fidelity and Fisher information metric, and point out that growth of Fisher information metric
 of thermofield double states perturbed by local operators corresponds to growth of commutator between excitation and external perturbation to Hamiltonian. This growth of commutators implies the initial excitation is scrambled.\\
 
 We consider time evolution of Fisher information metric of excited states. The inner product of $e^{-iH_{\lambda}t}Ae^{iH_{\lambda}t}|\psi\rangle$ and $e^{-iH_{\lambda+\delta\lambda}t}Ae^{iH_{\lambda+\delta\lambda}t}|\psi'\rangle$ is
 
  \be
 \langle\psi'|e^{-iH_{\lambda+\delta\lambda}t}Ae^{iH_{\lambda+\delta\lambda}t}e^{-iH_{\lambda}t}Ae^{iH_{\lambda}t}|\psi\rangle.\label{AAA}
  \ee\\
  The decay of this inner product can be understood as a butterfly effect. 
  When $|\psi_{\lambda}\rangle=|\psi_{\lambda+\delta\lambda}\rangle$, this inner product can be interpreted differently. 
  In this case, the quantity is called Polarization echo, and is given by
 
 \be
 \langle\psi|e^{iH_{\lambda+\delta\lambda}t}e^{-iH_{\lambda}t}Ae^{iH_{\lambda}t}e^{-iH_{\lambda+\delta\lambda}t}A|\psi\rangle.
 \ee\\
 When $|\psi\rangle$ is an eigenstate of $A$, Polarization echo can be used to measure the distance between
 a state $e^{iH_{\lambda}t}e^{-iH_{\lambda+\delta\lambda}t}|\psi\rangle$ and eigenspace of $A$ which $|\psi\rangle$ belongs to.\\
  
 \begin{figure}[ttt]
\begin{center}
\includegraphics[width=0.45\columnwidth]{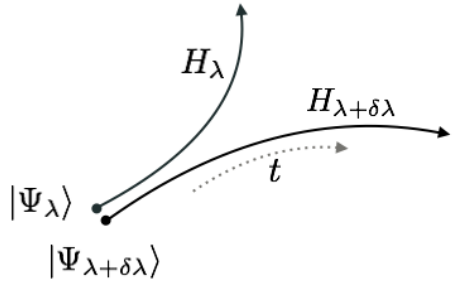}
\caption{Orbits of states with slightly different Hamiltonians. For chaotic theory or chaotic perturbation to the original Hamiltonian, the distance between states grows exponentially even if original states are the same or close to each other.
\label{fig:butterfly}}
\end{center}
\end{figure}

  Polarization echo and related quantities are studied and measured in various experiments, including
 Nuclear Magnetic Resonance, Microwave billiards, Elastic waves in metals, and so on.
 Fisher information metric is a useful tool in quantum estimation theory, in particular, its reciprocal gives
   lower bound of variance of an estimate of deterministic parameter in Cramer-Rao bound. Also,
    it is known that they can be used as diagnosis of non-Markovianity \cite{Breuer}\cite{Haikka}.
    Fisher information metric and fidelity are also used as order parameter for quantum phase transitions \cite{Quan}\cite{Zanardi}.\\
  
 Thermofield double state of a QFT is defined as a purification of thermal density matrix.
 Explicitly, it is defined by

\be
|\Psi_{\rm TFD}\rangle=\f{1}{\sqrt{Z(\beta)}}\sum_{i}e^{-\f{\beta E_{i}}{2}}|E_{i}\rangle_{L}\otimes|E_{i}\rangle_{R},
\ee\\
where $\beta$ is inverse temperature and $|E_{i}\rangle$ is eigenstate of QFT Hamiltonian.
The TFD state lives in tensor product of two identical QFT Hilbert spaces, and tracing out either side of Hilbert space
 gives the thermal density matrix of the system.
Time evolution of TFD state is given by

\ba
|\Psi_{\rm TFD}(t)\rangle&&=e^{-i(H_{L}+H_{R})t}|\Psi_{\rm TFD}\rangle\no&&
=\f{1}{\sqrt{Z(\beta)}}\sum_{i}e^{-(\f{\beta}{2}+2it)E_i}|E_{i}\rangle_{L}\otimes|E_{i}\rangle_{R},
\ea\\
where $Z(\beta)$ is thermal partition function.
We consider scalar excitation $W$ at $t=-t_w$ on TFD state, which lives in left QFT
 Hilbert space.
Then the perturbed TFD state at $t=0$ is given by

\be
|\Psi_{\rm TFD}(\lambda,W)\rangle=\f{1}{\sqrt{Z_W(\beta)}}e^{-iH^{\lambda}_{tot}t_w}We^{iH^{\lambda}_{tot}t_w}|\Psi_{\rm TFD}(\lambda)\rangle,
\ee\\
with $H_{tot}=H_{L}+H_{R}$. Tracing out right Hilbert space of this state gives correct excited thermal density
 matrix

\be
e^{-iH^{\lambda}t_w}We^{iH^{\lambda}t_w}\rho(\beta)e^{-iH^{\lambda}t_w}We^{iH^{\lambda}t_w}.
\ee\\
Let's consider deformation parametrized by coupling $\lambda$, of the QFT Hamiltonian. Then the inner product and the Fisher information metric of TFD state $|{\rm TFD}(t,\lambda)\rangle$ corresponding to (\ref{AAA}) with $|\psi\rangle\neq|\psi'\rangle$ is given by

\ba&&
|\la\Psi_{\rm TFD}(\lambda+\delta\lambda,W)|\Psi_{\rm TFD}(\lambda,W)\lb|
\no&&=1-G_{\lambda\lambda}(\delta\lambda)^2+O((\delta\lambda)^3).\ea\\
 We assume that the theory has time reversal symmetry. Then 
 
 \ba&&
|\la\Psi_{\rm TFD}(\lambda+\delta\lambda,W)|\Psi_{\rm TFD}(\lambda,W)\lb|
\no&&= \f{1}
{\sqrt{Z_W(\beta,\lambda)Z_W(\beta,\lambda+\delta\lambda)}}
|{\rm Tr}{\Big [}e^{-\f{\beta}{2}H_{L}^{\lambda+\delta\lambda}}W^{\lambda+\delta\lambda}(-t_w)W^{\lambda}(-t_w)e^{-\f{\beta}{2}H^{\lambda}_{L}}{\Big ]}|,\no
\ea\\
where we defined $W^\lambda(t)=e^{iH^{\lambda}t}We^{-iH^{\lambda}t}$.
The Fisher information metric can be decomposed into two parts,

\ba
G_{\lambda\lambda}=G_{\lambda\lambda}^{(W:0)}+G_{\lambda\lambda}^{(W:c)},\label{00}
\ea\\
where 

\ba
G_{\lambda\lambda}^{(W:0)}=&&\f{1}{2}\int_{0}^{\beta} \underset{t_1>\f{\beta}{2}>t_2}{dt_{1}dt_{2}}
{\rm Tr}{\Big [}\f{e^{-\beta H}}{Z_W(\beta,\lambda)}e^{Ht_1}Ve^{-Ht_1}e^{Ht_2}Ve^{-Ht_2}W(-t_w)^2{\Big ]}
\no&&
-\f{3}{8}(\int_{0}^{\beta}dt{\rm Tr}{\Big [}\f{e^{-\beta H}}{Z_W(\beta,\lambda)}e^{Ht}Ve^{-Ht}W(-t_w)^2{\Big ]})^2,\label{11}\ea\\
\ba
G_{\lambda\lambda}^{(W:c)}
=&&{\rm Re}{\Big [}~\f{1}{2}\int_{0}^{t_w}dt_{1}dt_{2}
{\rm Tr}{\Big [}\f{e^{-\beta H}}{Z_W(\beta,\lambda)}{\Big [}W(-t_w),~V(-t_1){\Big ]}\cdot{\Big [}W(-t_w),~V(-t_2){\Big ]}^{\dagger}{\Big ]}\no&&
+\f{i}{2}\int_{0}^{\f{\beta}{2}} dt_{E}\int_{0}^{t_w} dt
{\rm Tr}{\Big [}\f{e^{-\beta H}}{Z_W(\beta,\lambda)}e^{Ht_E}Ve^{-Ht_E}{\Big[}~{\Big [}W(-t_w),~V(-t){\Big]},~W(-t_w){\Big ]}{\Big ]}~{\Big ]}\no&&
-\f{1}{2}(\int_{0}^{t_w}dt{\rm Tr}{\Big [}\f{e^{-\beta H}}{Z_W(\beta,\lambda)}{\Big [}W(-t_w),~V(-t){\Big ]}~W(-t_w){\Big ]})^2.
\label{22}\ea\\
$G_{\lambda\lambda}^{(W:c)}$ is negligible when the theory is not chaotic, so that $[W(-t_w),~V(-t)]\approx 0$ holds.
Therefore, the growth or decay of $G_{\lambda\lambda}^{(W:c)}$ implies scrambling is taking place. \\

Let us consider large N limit.
The relaxation time $t_r$ of a thermal QFT can be determined from the behavior of two point function 

\be
\langle W(t)W(0)\rangle\sim \mathcal{O}(e^{-\f{t}{t_r}}).
\ee\\
In holographic CFT, $t_r$ is given by $\beta$, which is significantly small compare to scrambling time $t_s=\f{\beta}{2\pi}~{\rm log}N^2$. We consider particular class of large N QFT with such hierarchy.

In \cite{Maldacena:2015waa}, it was shown that CFT dual to Einstein gravity has largest Lyapunov exponent, 
among large N theories with such hierarchy.

Under these assumptions, 

\be
G_{\lambda\lambda}^{(W:0)}\approx \f{1}{2}\int_{0}^{\beta} \underset{t_1>\f{\beta}{2}>t^2}{dt_{1}dt_{2}}
{\rm Tr}\big[\f{e^{-\beta H}}{Z(\beta,\lambda)}e^{Ht_1}Ve^{-Ht_1}e^{Ht_2}Ve^{-Ht_2}\big].
\ee\\
Therefore, $G_{\lambda\lambda}^{(W:0)}$ coincides with unexcited Fisher information metric.
In particular, $G_{\lambda\lambda}^{(W:0)}$ is independent of excitation $W$ and is time independent.
This fact implies that growth of Fisher information metric comes purely from scrambling in large N theory.  
\\

We can also consider similar quantity, corresponding to (\ref{AAA}) with $|\psi\rangle=|\psi'\rangle$ , which purely contains chaotic terms.
We consider the inner product of two states

\be
|\Psi_{\rm TFD}(\lambda,W)\rangle=\f{1}{\sqrt{Z_W(\beta)}}e^{-iH^{\lambda}_{tot}t_w}We^{iH^{\lambda}_{tot}t_w}|\Psi_{\rm TFD}(\lambda)\rangle,
\ee
\be
|\Psi_{\rm TFD}(\lambda,\delta\lambda,W)\rangle=\f{1}{\sqrt{Z_W(\beta,\delta\lambda)}}e^{-iH^{\lambda+\delta\lambda}_{tot}t_w}We^{iH^{\lambda+\delta\lambda}_{tot}t_w}|\Psi_{\rm TFD}(\lambda)\rangle,
\ee\\
where $Z_W(\beta,\delta\lambda)$ is the normalization constant. 
Infinitesimal part of the inner product is
 defined by

\ba&&
|\langle\Psi_{\rm TFD}(\lambda,W)|\Psi_{\rm TFD}(\lambda,\delta\lambda,W)\rangle|
\no&&=1-G_{\lambda\lambda}^{S}(\delta\lambda)^2+O((\delta\lambda)^3)
\no&&= \f{1}
{\sqrt{Z_W(\beta,\lambda)Z_W(\beta,\delta\lambda)}}
|{\rm Tr}{\Big [}e^{-\beta H_{L}^{\lambda}}W^{\lambda+\delta\lambda}(-t_w)W^{\lambda}(-t_w){\Big ]}|
.\ea\\
$G_{\lambda\lambda}^{S}$ can be explicitly written as

\ba
G_{\lambda\lambda}^{S}
=&&{\rm Re}{\Big [}~\f{1}{2}\int_{0}^{t_w}dt_{1}dt_{2}
{\rm Tr}{\Big [}\f{e^{-\beta H}}{Z_W(\beta,\lambda)}{\Big [}W(-t_w),~V(-t_1){\Big ]}\cdot{\Big [}W(-t_w),~V(-t_2){\Big ]}^{\dagger}{\Big ]}~{\Big ]}\no&&
-\f{1}{2}(\int_{0}^{t_w}dt{\rm Tr}{\Big [}\f{e^{-\beta H}}{Z_W(\beta,\lambda)}{\Big [}W(-t_w),~V(-t){\Big ]}~W(-t_w){\Big ]})^2.
\label{33}\ea\\
Compared to (\ref{00}), this quantity does not include constant divergent terms, but only contains terms proportional to
 commutators of operators. Therefore, we can use this quantity to measure the growth of commutators, therefore scrambling.

\section{Holographic caluculation}

In this section, we will show Fisher information metric grows rapidly in holographic systems, using proposed
 holographic dual of Fisher information metric.\\

The metric of $d+1$ eternal black hole is given by

\be
ds^2=-f(r)dt^2+f(r)^{-1}dr^2+r^2d\Omega_{S^{d-1}}^2,
\ee\\
where $f(r)$ is given by $r^2-R^2$ for $d=2$ with $R^2=8G_{N}M$,
 and $r^2+1-a^2/r^{d-2}$ for higher dimensions with $a^2=\f{G_N M}{{\rm Vol}(S^{d-1})}$.
$R$ is radius of black hole horizon, and AdS radius is set to 1.
The inverse Hawking temperature $\beta$ of this black hole is given by $\beta=\frac{4\pi}{f'(R)}$.
\begin{figure}[ttt]
\begin{center}
\includegraphics[width=0.45\columnwidth]{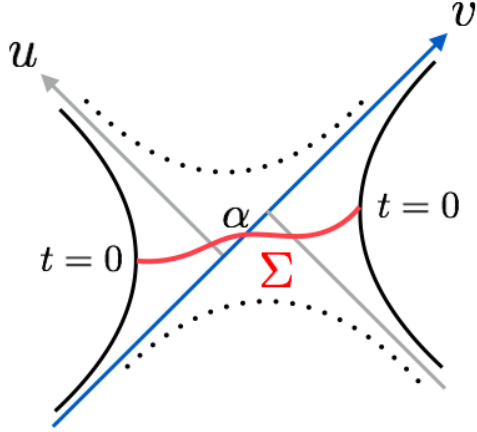}
\caption{A diagram of the shock wave geometry. Black line are boundaries, dashed lines are singularities, and a blue line is the shock wave.
Red surface connecting two boundaries is the extremal volume surface $\Sigma$. For 2d CFT, this surface can be understood as configuration of world sheet of a string ending on both boundaries. 
\label{fig:shock}}
\end{center}
\end{figure}
In the Kruskal coordinate,  the metric is

\be
ds^2=-\f{f(r)}{f'(R)^2}e^{-f'(R)r_{*}(r)}dudv+r^2d\Omega_{S^{d-1}}^2,
\ee\\
where $u$ and $v$ are given by

\ba&&
u^2=e^{f'(R)(r_{*}(r)-t)}\no&&
v^2=e^{f'(R)(r_{*}(r)+t)}.
\ea\\
and $r_{*}(r)$ is defined to satisfy $dr_{*}=f(r)^{-1}dr$ and $r_{*}(0)=r_{*}(\infty)=0$.
The state dual to eternal black hole at $t=0$ is thermofield double state \cite{Maldacena:2001kr}\cite{Hartman:2013qma}.
We consider a few particles with total energy $E$ are thrown homogeneously into eternal black hole at $t=-t_{w}$ from left boundary. 
The proper energy of the particles at $t=0$ slice is $E_p\approx \frac{E}{R}e^{\frac{2\pi}{\beta}t_w}$. So when
  $t_w$ is sufficiently large, the particle becomes shock wave, and the geometry can be approximated by
   shock wave geometry, which is given by glueing two black holes with mass $M$ and mass $M+E$. It is given by $u_L=u_R$ and $v_L=v_R+\alpha$, where $\alpha$ is defined by

\be
\alpha=\f{E}{u_w}\f{dR}{dM}C(R,R)
\ee\\
with black hole mass $M$ and $C(r,R)=\f{e^{f'(R)r_{*}(r)}}{r-R}$.
In 2d, $\alpha$ is given by $\alpha=\f{E}{4M}e^{\f{2\pi}{\beta}t_w}$.
In higher dimensions, $\alpha$ is proportional to $G_{N}\cdot E\cdot e^{\f{2\pi}{\beta}t_w}$.

According to the proposal in \cite{MIyaji:2015mia}, inner product of TFD states is holographically given by
 the volume of codimension 1 extremal volume surface $\Sigma$, which connects $t=0$ slices at two boundaries
 of spacetimes. This is given as
 
 \be
 |\la\Psi_{TFD}(t, \lambda+\delta\lambda)|\Psi_{TFD}(t,\lambda)\lb|=e^{-n_{d}(\delta \lambda)^2{\rm Vol}(\Sigma)},
  \ee\\
  where $n_d$ is $\mathcal{O}(1)$ constant.
  The Fisher information metric is then
 
 \be
 G_{\lambda\lambda}=n_{d}{\rm Vol}(\Sigma).
   \ee\\
 The idea of this proposal stems from Janus geometry \cite{Janus}
   and the proposed duality between AdS/CFT and MERA \cite{Swingle:2009bg}\cite{Miyaji:2014mca}. 
   We note that, in \cite{Stanfords:2014cas}, the volume of maximal volume surface in the bulk
    was conjectured to be equal to the complexity of the corresponding boundary state.    
    
Let's insert a homogeneous scalar operator W on left boundary at $t=-t_w$
and consider the perturbed TFD state at $t=0$. The resulting holographic geometry is given by 
the shock wave geometry. The volume of the surface connecting two boundaries at $t=0$ is

\be
{\rm Vol}({\rm \Sigma})=\int dtd\Omega_{d-1}\sqrt{(-f+\dot{r}^2f^{-1})r^{2(d-1)}}.
\ee\\
When the surface has extremal volume, the conserved quantity along the surface is

\be
s_d=\f{f(r)r^{d-1}}{\sqrt{-f(r)+\dot{r}^2f(r)^{-1}}}.
\ee\\
Therefore we conclude that the volume of extremal volume surface is

\be
{\rm Vol}({\rm \Sigma})={\rm Vol}(S^{d-1})\int dr r^{d-1}\sqrt{f(r)^{-1}(1-\f{1}{1+\f{f(r)r^{2(d-1)}}{s_d^2}})}.
\ee\\
Let us define $r_0$ by $\dot{r}(r)=0$ and corresponding time as $t_0$.
Then we can express $\alpha$ in terms of $s_d$ as

\be
{\rm log}\f{\alpha}{2}=
r_{*}(r_0)+t_0+\int_{r_0}^{R} dt\f{1}{|f(r)|\sqrt{1+\f{f(r)r^{2(d-1)}}{s_d^2}}}(1-\sqrt{1+\f{f(r)r^{2(d-1)}}{s_d^2}}).
\ee\\

\begin{figure}[ttt]
\begin{center}
\includegraphics[width=0.65\columnwidth]{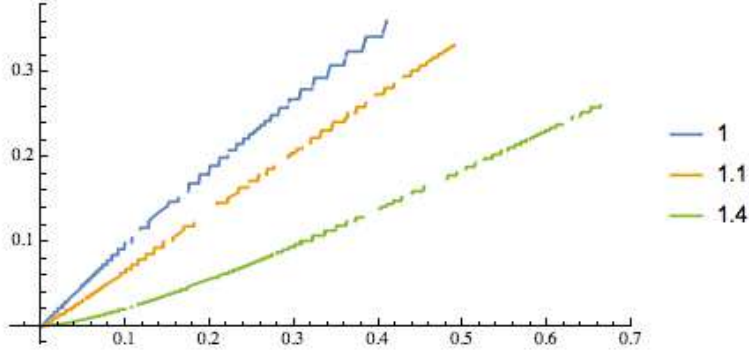}
\caption{Numerical plot for growth of information metric versus $\alpha=\f{E}{4M}e^{\f{2\pi}{\beta}t_w}$ in $AdS_3/CFT_2$. The numbers in the right are radiuses of BH horizon $R=\f{2\pi}{\beta}$.The vertical ax is for $\f{2\Delta{\rm Vol}}{R^2}$, and horizontal ax is for $\alpha$. We can confirm linear growth of information
 $\Delta{\rm Vol}=f_1\times \alpha+\mathcal{O}(\alpha^2)$ for small $\alpha$. 
\label{fig:123}}
\end{center}
\end{figure}

The numerical result is shown in fig (\ref{fig:123}), and we can conclude that 

\be
G_{\lambda\lambda}^{W:c}=f_1\cdot \alpha+\mathcal{O}(\alpha^2).
\ee\\
where $f_{1}$ is $N$ independent and positive, but has dependence on the temperature.
For $d=2$ at $R=1,~1.1,~1.4$, we get $f_1\approx 0.9,~0.7,~0.2$, respectively.
This implies that the information metric for perturbed TFD states for external fields
 grows rapidly, proportional to $\alpha$. 
 
 Although approximation by classical gravity is no longer applicable when $\alpha>>1$, we can estimate
 the behavior of inner product by sticking to classical computation. Numerical result in $CFT_2$ is

\be
\f{|\langle\Psi_{TFD}(\lambda+\delta\lambda,~W)|\Psi_{TFD}(\lambda,~W)\rangle|}{|\langle\Psi_{TFD}(\lambda+\delta\lambda)|\Psi_{TFD}(\lambda)\rangle|}\underset{\alpha >>1}{\sim} (\f{1}{\alpha})^{\f{2\pi^2n_d}{\beta^2}(\delta\lambda)^2}\propto e^{-\f{4\pi^3n_d}{\beta^3}t_w(\delta\lambda)^2}.
\ee\\
So we can observe the exponential decay of inner product.\\

\section{Wilson loop}
We can study two point functions of Wilson loops on different boundaries, using holographic prescription \cite{Rey:1998ik} 
\cite{Maldacena:1998im}. We assume these loops are put on the great circles of $S^{d-1}$ on different boundaries. It is expected that such
 two point functions should decay exponentially in time. In 2d CFT, the two point function is given by $\langle W_LW_R\rangle\sim e^{-(S_{NG}-reg)}$, 
using extremal value of Numbu-Goto action,

\ba
S_{NG}&&=\f{1}{2\pi\alpha'}\int dtd\Omega_{1}\sqrt{(-f+\dot{r}^2f^{-1})r^{2}}\no&&
=\f{1}{2\pi\alpha'}{\rm Vol}(S^1)\int dr ~r\sqrt{f(r)^{-1}(1-\f{1}{1+\f{f(r)r^{2}}{s_2^2}})}.
\ea\\
The result is same as the calculation of volume of extremal codimension 1 surface, up to constant factor.
Therefore, two point function of Wilson loops decays as $\sim f_0-f_1\times\alpha+\cdots$ for small $\alpha$, as expected.  

\section{Conclusion}
In this article, we estimated the time evolution of information metric and inner product, of thermofield double state
 perturbed by an operator $W$ at $t=-t_w$. In large N theories, information metric can be separated into two terms,
  one is unperturbed information metric, and the other is proportional to commutator of $V$ and $W(-t_w)$.
  Rapid growth of the later term indicates scrambling of $W$, and sensitivity of time evolution of 
  thermofield double state or thermal mixed states to external environment or internal imperfections. Indeed, we calculated the information metric and confirmed
   its rapid growth proportional to $e^{\f{2\pi t}{\beta}}$.
   This implies that decay of OTO correlator can be understood by growth of distance between different thermal states.
   We also studied two point function of Wilson loops and confirmed that it decays exponentially.\\
  
  It is interesting to calculate Fisher information metrics and inner products for pure states or reduced density matrices on subregions. Our study focused only on small perturbations to original Hamiltonian, but we can also
 study Loschmidt echo for strong perturbations. In that case, the decay rate is expected to be independent of strength of perturbations, and the rate can be related to classical Lyapunov exponent of the system. 
 Furthermore, it would be very interesting to explore physical quantities more, which capture scrambling, and have simple gravity correspondents.
  
\section*{Acknowledgements}

  We are grateful to Tadashi Takayanagi and Pawel Caputa for fruitful discussions and comments. We also
  appreciate Beni Yoshida, Tokiro Numasawa and Satoshi Iso for useful discussions. We thank the workshop "Quantum Information in String Theory and Many-body Systems" at Yukawa Institute for Theoretical Physics, Kyoto University. MM is supported by JSPS fellowship.

\end{document}